\documentclass[aps,prl,twocolumn,superscriptaddress,showpacs]{revtex4}
\usepackage{graphicx}
\usepackage{bm}
\usepackage{amsmath}

\begin{document}


\title{Direct observation of quantum coherence in single-molecule magnets}


\author{C. Schlegel}
\affiliation{1. Physikalisches Institut, Universit\"at Stuttgart, Pfaffenwaldring 57, 70550 Stuttgart, Germany}
\author{J. van Slageren}
\affiliation{1. Physikalisches Institut, Universit\"at Stuttgart, Pfaffenwaldring 57, 70550 Stuttgart, Germany}
\affiliation{School of Chemistry, University of Nottingham, Nottingham NG7 2RD, United Kingdom} \email{joris.van.slageren@nottingham.ac.uk}
\author{M. Manoli}
\author{E.K. Brechin}
\affiliation{School of Chemistry, University of Edinburgh, West Mains Road, Edinburgh, EH9 3JJ, United Kingdom}
\author{M. Dressel}
\affiliation{1. Physikalisches Institut, Universit\"at Stuttgart, Pfaffenwaldring 57, 70550 Stuttgart, Germany}


\date{\today}

\begin{abstract}
Direct evidence of quantum coherence in a single-molecule magnet in frozen solution is reported with coherence times as long as $T_2 = 630\pm30$ ns. We can strongly increase the coherence time by modifying the matrix in which the single-molecule magnets are embedded. The electron spins are coupled to the proton nuclear spins of both the molecule itself and interestingly, also to those of the solvent. The clear observation of Rabi oscillations indicates that we can manipulate the spin coherently, an essential prerequisite for performing quantum computations.
\end{abstract}

\pacs{03.67.-a, 75.30.Gw, 75.50.Xx, 76.30.-v}

\maketitle

\begin{figure}
\includegraphics[width=8cm]{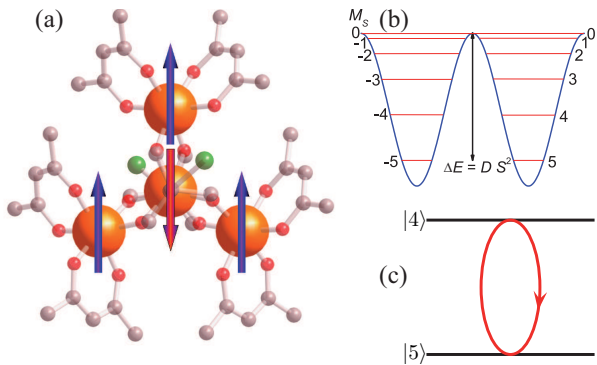}
\caption{\label{Fig.1}{FIG. 1 (color online). (a) Crystal structure of Fe$_4$ with iron ions depicted as large orange spheres. The arrows denote the relative orientations of the magnetic moments of each iron ion in the \emph{S} = 5 ground state. (b) Double-well potential energy diagram depicting the energy barrier between spin up and spin down. (c) Resonant photon-spin interaction (Rabi cycle) between magnetic sublevels.}}
\end{figure}
The key concept in quantum information processing is that a quantum bit (qubit) may be not just 0 or 1, as in ordinary computer bits, but an arbitrary superposition of 0 and 1. This means that any two-level system, that can be put into a superposition state, is a qubit candidate \cite{Sto04}. The required superposition state is created by electromagnetic radiation pulses with a frequency corresponding to the energy splitting between the two levels (Fig. \ref{Fig.1}c). The contribution of each of the two levels to the superposition state has a cyclic dependence on the pulse length, leading to so-called Rabi oscillations~\cite{Sto04}. The observation of such oscillations is a proof-of-principle for the viability of performing quantum computations with a particular system. Quantum computers will probably not be
realized from single atoms but will most likely utilize solid state devices,
such as superconducting junctions, semiconductor structures, or molecular magnets \cite{Sto04,Leu01}. For these large systems the quantum coherence decays fast, which drastically shortens the time available for quantum computation.

 Molecular magnets have been considered as qubits because they can be easily organized into large-scale ordered arrays by surface self-assembly \cite{Gom06}, and because they possess excited electronic-spin states required for two-qubit gate operations \cite{Car07,Aff06}. Single-molecule magnets (SMMs) are exchange-coupled clusters with high-spin ground states~\cite{Gat06}. The Ising-type anisotropy creates an energy barrier toward magnetization relaxation [Fig. \ref{Fig.1}(b)], and many fascinating quantum phenomena  have been observed in these systems, such as quantum tunnelling of the magnetization and quantum phase interference \cite{Gat06}. The large splitting of the two lowest states of SMMs in zero field (in principle) allows performing coherent spin-manipulations \emph{without} external magnetic field, which simplifies any practical implementation. SMMs have also been proposed for the implementation of Grover's algorithm \cite{Leu01} allowing numbers between 0 and 2$^{2S-2}$ to be stored in a single molecule.

The long coherence time is a crucial first step towards successful implementation of SMMs as qubits \cite{Sto04}. Therefore, recent years have seen a great deal of activity in trying to determine the quantum coherence times in SMMs, which was estimated to be of the order of 10 ns \cite{Wer05,Del04,Hil03,DeL08}. In several cases, energy gaps between superposition states have been reported that are larger than the expected decoherence energy scale \cite{Hil03, Lui00, Wal05}. The phase memory or decoherence time of SMMs remains unresolved, although magnetization detected ESR studies using pulsed microwave irradiation gave some indication of the spin dynamics \cite{Bah07, Bal08}. In the low-spin system ferritin, on the other hand, there is evidence of quantum coherence \cite{Gid95}. Using pulsed electron-spin resonance (ESR) \cite{Sch01}, spin-spin relaxation or coherence times (denoted $T_2$) were determined in several molecular magnets and metalloproteins with $S=\frac{1}{2}$ ground states.
In iron-sulfur clusters, for instance, $T_2$ is several hundreds of nanoseconds \cite{Dil99,She91,Gui95}, while in the Cr$_7$Ni and Cr$_7$Mn antiferromagnetic rings, $T_2\approx 400$~ns at 4.5 K, increasing to 3.8~$\mu$s for deuterated Cr$_7$Ni at 1.8~K~\cite{Ard07}; and $T_2= 2.6~\mu$s for an antiferromagnetic iron(III) trimer \cite{Mit08}. Recently, quantum oscillations were observed in the molecular magnet V$_{15}$~\cite{Ber08}. Bertaina et al. also studied Rabi-oscillations in a system with large angular momentum \cite{Ber07}.

Here we show the first direct experimental evidence for long-lasting quantum coherence and quantum oscillations in a SMM, by using pulsed W-band (94.3 GHz) ESR spectroscopy. We investigated the Fe$_4$ complex [Fe$^{\text{III}}_4$(acac)$_6$(Br-mp)$_2$] (Fig. \ref{Fig.1}a), where acac is acetyl acetonate or 2,4-pentanedionate, and Br-mp$^{3-}$ is the anion of 2-(bromomethyl)-2-(hydroxymethyl)-1,3-propanediol (Br-mpH$_3$) \cite{Man08}. This molecule was chosen because the zero-field splitting is such that the ground state magnetic resonance transition is close to zero field in our spectrometer, in contrast to the case of other thoroughly investigated and chemically similar Fe$_4$ clusters \cite{Acc06}. In this molecule the central iron(III) spin is coupled antiferromagnetically to the peripheral iron(III) spins resulting in an \emph{S} = 5 molecular spin ground state. The $M_S$ states of this multiplet are split with the $M_S = \pm S$ at lowest energy (Ising type anisotropy). This anisotropy leads to a splitting between the lowest two spin-microstates of 92.4 GHz, and the corresponding ESR transition occurs conveniently close to zero applied field in our Bruker ELEXSYS E680 W-band ESR spectrometer \cite{Man08}. The material was diluted into a frozen solvent matrix to limit decoherence due to intermolecular magnetic-dipolar interactions, which is the main decoherence pathway in crystalline samples \cite{Mor06}. For lower concentrations lower than the used 0.5 mg/mL, experiments did not show a significant increase in spin relaxation times. In the following, we report the determined spin-lattice relaxation time ($T_1$) and the phase coherence time ($T_2$), the coupling between electron and nuclear spins and finally transient nutation experiments that show the occurrence of Rabi-oscillations.

\begin{figure*}
\includegraphics[width=15cm]{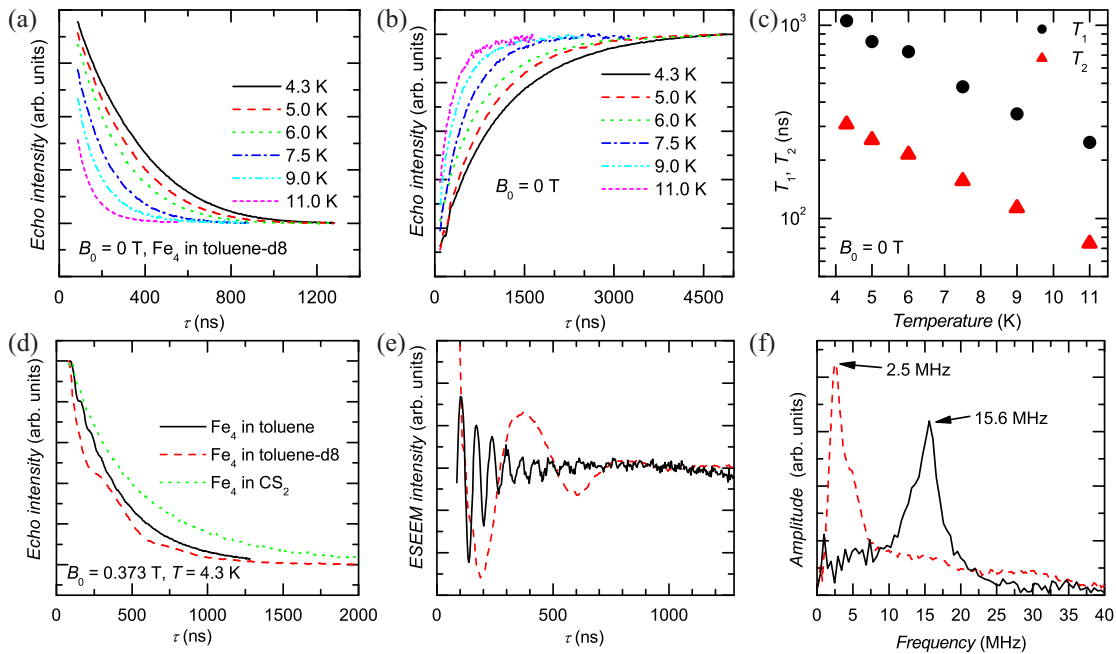}
\caption{\label{Fig.2}{(color online) (a) Normalized Hahn-echo intensity of Fe$_4$ in toluene at 0 T at different temperatures; the data were rescaled so that the intensities at zero delay time would match. (b) Rescaled echo intensity at 0 T after an inversion recovery sequence of Fe$_4$ in toluene at different temperatures as indicated in the Figure. (c) Temperature-dependent spin-spin ($T_2$) and spin-lattice ($T_1$) relaxation times of Fe$_4$ in toluene obtained from the fit of data from (a) and (b), respectively. (d) Echo decay for Fe$_4$ in toluene (black, solid), toluene-d$_8$ (red, dashed) and CS$_2$ (green, dotted) at 0.373 T and 4.3 K. (e) ESEEM modulation after subtraction of the exponential decay from (d). (f) ESEEM frequencies from Fourier transformation of data in (e).}}
\end{figure*}

Fig. \ref{Fig.2}(a) shows the Hahn echo intensity of the Fe$_4$ complex in toluene as a function of delay time $\tau$ ($\pi/2-\tau-\pi-\tau-echo$, $\pi/2$ pulse length is 14 ns), at zero external field for \emph{T} = 4.3 - 11.0 K. At all temperatures, the decay is monoexponential where the time constant is the coherence time $T_2$. At \emph{T} = 4.3 K, $T_2 = 307\pm20$ ns and it decreases strongly with increasing temperature. The clear observation of a Hahn echo unambiguously proves that quantum coherence in Fe$_4$ is much longer than previously estimated for SMMs.

The spin-lattice relaxation time $T_1$ [Fig. \ref{Fig.2}(b)] was determined by using an inversion recovery sequence ($\pi-\tau-\pi/2-\tau_{\text{fixed}}-\pi-\tau_{\text{fixed}}-echo$). At \emph{T} = 4.3 K, $T_1 = 1056\pm20$ ns [Fig. \ref{Fig.2}(c)] and it decreases with increasing temperature. The strong temperature dependence of $T_1$ evidences that spin-lattice relaxation must occur through a two-phonon process, because the direct process is expected to be little dependent on temperature \cite{Abr86}. We cannot clearly distinguish between Raman and Orbach processes, due to the limited temperature range accessible. However, in exchange-coupled clusters the spin-lattice relaxation mechanism is often an Orbach process where an excited spin state functions as the intermediate state \cite{Ben90}. A fit to the Orbach formula yields an energy gap of the order of $\Delta = 5$ cm$^{-1}$, which is clearly within the ground multiplet.

The (super)hyperfine coupling between the nuclear and electron spins is known to be the major decoherence path in both molecular and nanostructured systems \cite{Sto04, Ard07,Pro00}. To investigate the coupling to the nuclear spin bath of the matrix, we investigated Fe$_4$ samples in three different solvents: normal and fully deuterated toluene, and CS$_2$. The first of these solvents contains a large number of protons to which the electron spin can couple. The second only contains deuterium atoms, which couple much more weakly, while in the third solvent nuclear spins are completely absent. Hahn echo measurements at an applied external field of $B_0$ = 0.373 T exhibit a clear modulation of the echo superimposed on the exponential decay [Fig. \ref{Fig.2}(d)] for samples in toluene or toluene-d$_8$. This modulation is due to the coupling of the nuclear spin to the electron spin (electron spin echo envelope modulation \cite{Sch01}, ESEEM). After subtraction of the background exponential decay, the dominant ESEEM frequencies can be observed both in time domain [Fig. \ref{Fig.2}(e)] and after Fourier transformation in the frequency domain [Fig. \ref{Fig.2}(f)]. The dominant ESEEM frequency of Fe$_4$ in toluene, $15.6\pm0.3$ MHz, is the same as the free Larmor frequency of protons at 0.373T (15.9 MHz). For Fe$_4$ in toluene-d$_8$ the dominant frequency is $2.5\pm0.2$ MHz, which is the free Larmor frequency of deuterium atoms (2.4 MHz). Interestingly, this indicates that we observe the coupling of the electron spin to the nuclear spin of the solvent rather than to the 58 protons of the Fe$_4$ molecule itself. Accordingly, no clear ESEEM was observed for Fe$_4$ in the CS$_2$ solvent, which has no significant amounts of nonzero nuclear spins. In toluene-d$_8$, $T_2$ is virtually the same ($279\pm20$ ns at 0T) as for normal toluene, which shows that decreasing the coupling to the nuclear spin bath does not increase coherence times, in contrast to what was observed in the measurements by Ardavan et al. \cite{Ard07}. However, for Fe$_4$ in CS$_2$, the coherence time $T_2$ increases dramatically to $527\pm20$ ns at 0T. This observation leads to two very important conclusions. First, the coupling to the Fe$_4$ nuclear spins is not the primary decoherence pathway, and, second, the solvent nuclear spins must be removed entirely to suppress decoherence. This increase of $T_2$ in CS$_2$ is accompanied by an increase in $T_1$, which was determined to be $T_1=943$ ns at 4.3 K and 0 T, which shows there is no simple relation between $T_1$ and $T_2$.

\begin{figure}
\includegraphics[width=7cm]{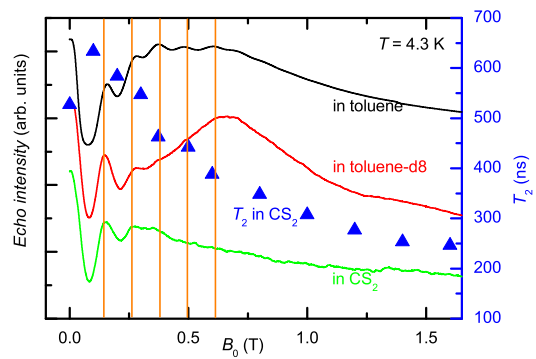}%
\caption{\label{Fig.3}{(color online) Normalized echo intensity recorded as a function of field for Fe$_4$ in toluene (top, black trace), in toluene-d$_8$ (middle, red trace) and CS$_2$ (bottom, green trace). Fields at which nuclear modulation maxima for the coupling to protons are expected are indicated with vertical orange lines. $T_2$ was determined at different fields in CS$_2$ (blue triangles, right-hand scale).}}
\end{figure}

\begin{figure}
\includegraphics[width=7cm]{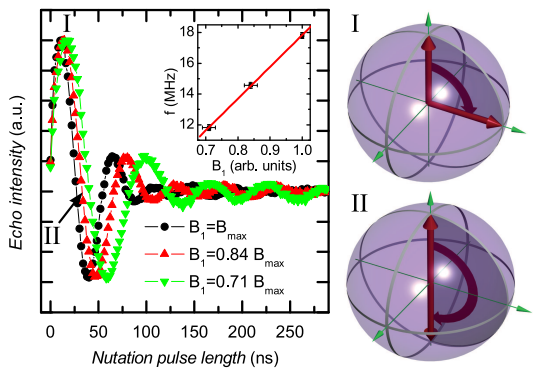}%
\caption{\label{Fig.4}{(color online) Rabi oscillations obtained by recording the echo intensity as a function of nutation pulse length on Fe$_4$ in CS$_2$ at 0 T and different driving field strengths $B_1$. For two positions in the Rabi cycle the corresponding path on the Bloch sphere is shown on the right.}}
\end{figure}

We performed echo-detected ESR measurements (Fig. \ref{Fig.3}), in which the echo intensity after the Hahn echo sequence ($\tau_{\text{fixed}} = 185$ ns) was recorded as a function of of static magnetic field $B_0$ for samples of Fe$_4$ in toluene, toluene-d$_8$ and CS$_2$. We see echo intensity over the whole field range from $0-2$ T, which is expected because molecules with different orientations with respect to the external magnetic field $B_0$ are excited at different fields over the entire studied field range. For Fe$_4$ in toluene we observe equally spaced modulations of the echo intensity between 0 and 0.7 T. Theory predicts maxima of this modulation at $B_n=2\pi n/\gamma\tau$ (\emph{n} = 0, 1, 2, ...), where $\gamma$ is the gyromagnetic ratio of the nucleus and $\tau$ is the pulse separation,\cite{Sch01} and the field positions of the observed maxima are consistent with calculated field values for $^1$H nuclei (Fig. \ref{Fig.3}). Surprisingly, we see the first two maxima of the echo modulation due to $^1$H nuclei also for Fe$_4$ in toluene-d$_8$ and CS$_2$, which shows that the electron spin is also coupled to intramolecular protons. The decrease in intensity for 0.2 T $< B_0 <$ 0.65 T in the spectrum of the toluene-d$_8$ sample agrees with the expected echo intensity minimum due to coupling to $^2$H nuclei at 0.41 T. At higher fields the modulation disappears, because the nuclear modulation depth is proportional to $B_0^{-2}$ \cite{Sch01}. The echo intensity decreases with field for $B_0 > 0.8$ T, despite the fact that one expects about the same number of molecules to be excited at all fields. This intensity decline is accompanied by a decrease in $T_2$ for $B_0 > 0.1$ T (Fig. \ref{Fig.3}), meaning that less spins can be refocussed after the fixed delay time of 185 ns, with concurrent smaller echo intensity. At $B_0 = 0.1$ T, $T_2$ reaches its maximum value of $T_2 = 630\pm30$ ns for Fe$_4$ in CS$_2$.\\
Finally, we have performed transient nutation experiments which correspond to generating arbitrary superposition states of the SMM. In the nutation experiment, the electron spin is rotated by an arbitrary angle, after which the spins are refocussed with a $\pi$ pulse and the echo intensity is detected \cite{Sch01}. Fig. \ref{Fig.4} displays the echo intensity as a function of the duration of the first pulse, showing clear intensity oscillations with a frequency of $17.6\pm0.5$ MHz at zero external field and maximum microwave power, i.e. largest $B_1$ field. These oscillations are coherent electron spin oscillations, i.e. they are the first demonstration of Rabi oscillations in SMMs. This interpretation is supported by the fact that the Rabi frequency depends linearly on the microwave power (Fig. \ref{Fig.4}) \cite{Sch01}. It should be noted that the oscillations cannot be due to ESEEM-type nuclear modulation because they occur at zero field. The decay of the Rabi oscillations with delay time is highly non-monoexponential for two reasons. First, the presence of a distribution in zero-field splitting parameters (\emph{D}-strain) can be expected, and second, there will be a distribution in orientation of the molecule with respect to the microwave field. Therefore, molecules with slightly different Rabi-frequencies will be excited, which leads to a faster decay of the observed Rabi oscillations than expected on the basis of the coherence time alone.\\
In conclusion, we have demonstrated that sizable quantum coherence times and coherent spin manipulations are possible in single-molecule magnets. Much longer quantum coherence times in the microsecond range can be predicted at lower temperatures based on the strong dependence of $T_2$ on temperature. For instance, the extrapolation of $T_2$ in Fig. \ref{Fig.2}(c) yields a coherence time of $T_2 \approx 750$ ns at 0 K, and even larger low-temperature $T_2$ values are expected for Fe$_4$ in CS$_2$. In addition, coherence times can be significantly improved by careful tailoring of the SMM and its surroundings and may reach values that are suitable for practical qubit implementation.\\
We gratefully acknowledge Professor M. Mehring, Professor G. Denninger, Dr. H.J. K\"ummerer and Dr. B. Naydenov for useful discussions and experimental assistance, and the DFG, EPSRC and Leverhulme Trust for funding.
\bibliography{Fe4}

\end{document}